\documentclass[aps,amsfonts,prd,twocolumn]{revtex4}
\usepackage{amsfonts}
\usepackage{amsmath}
\usepackage{amssymb}
\usepackage{epsfig}
\usepackage{graphicx}

\newcommand{\ar}{\arrowvert}
\newcommand{\bs}{\boldsymbol}
\newcommand{\mb}{{\mbox {\boldmath  $\alpha$ \unboldmath}}} 
\newcommand{\ra}{\rangle}
\newcommand{\la}{\langle}
\newcommand{\da}{\dagger}

\newcommand{\cd}{\! \cdot \!}
\newcommand{\be}{\begin{equation}}
\newcommand{\ee}{\end{equation}}
\newcommand{\ba}{\begin{eqnarray}}
\newcommand{\ea}{\end{eqnarray}}

\begin{document}
\title{The BES {\boldmath $ f_0(1810)$}:  a new glueball 
candidate}
\author{Pedro Bicudo}
\affiliation{Departamento de F\'{\i}sica and CFTP, Instituto Superior Tecnico, 
Lisboa, Portugal}
\author{Stephen R. Cotanch}
\affiliation{Department of Physics, North Carolina State University, Raleigh, 
NC 27695, USA}
\author{Felipe J. Llanes-Estrada}
\affiliation{Departamento de F\'{\i}sica Te\'orica I, Universidad Complutense de 
Madrid, 28040 Madrid, Spain}
\author{David G. Robertson}
\affiliation{Department of Physics and Astronomy, Otterbein College, Westerville, OH 43081, USA}


\begin{abstract}
We analyze the $f_0(1810)$ state recently observed by the BES
collaboration via radiative $J/\psi$ decay to a resonant $\phi\omega$
spectrum and confront it with DM2 data and glueball theory. The DM2
group only measured $\omega\omega$ decays and reported a pseudoscalar
but no scalar resonance in this mass region. A rescattering mechanism
from the open flavored $ K\bar{K}$ decay channel is considered to
explain why the resonance is only seen in the flavor asymmetric
$\omega \phi$ branch along with a discussion of positive $C$ parity
charmonia decays to strengthen the case for preferred open flavor
glueball decays. We also calculate the total glueball decay width to
be roughly 100 MeV, in agreement with the narrow, newly found $f_0$,
and smaller than the expected estimate of 200-400 MeV.  We conclude
that this discovered scalar hadron is a solid glueball candidate and
deserves further experimental investigation, especially in the $K
\bar{K}$ channel. Finally we comment on other, but less likely,
possible assignments for this state.
\end{abstract}
\maketitle

\section{Introduction}

Significant in the recent wave of particle discovery, the BES
collaboration has just reported \cite{bes} a scalar hadron with mass
about 1812 MeV and width of 105(20) MeV.  This $f_0(1810)$ state
appeared as a 95 event enhancement in the $\omega\phi$ spectrum
accompanied by a radiative photon from $5.8\times 10^7$ $J/\psi$
decays. This paper considers several interpretations for this state
and focuses on the most exciting assignment, the long-sought scalar
glueball.

Glueballs have been predicted in lattice gauge calculations
\cite{Morningstar:1999rf} and many body theory 
\cite{Szczepaniak:1995cw,Llanes-Estrada:2000jw} with both approaches agreeing
the ground state has quantum numbers $0^{++}$ and mass in the range
1700 to 1800 MeV.  Scalar hadrons between 1 and 2 GeV, predominantly
the $f_0(1370)$, $f_0(1500)$ and $f_0(1710)$, have been scrutinized
for glueball wavefunction components in numerous studies
\cite{Giacosa:2005zu,Anisovich:2001ig}; however, firmly identifying
gluonic degrees of freedom remains elusive \cite{Bugg:2004xu}.  In
this paper we dispel several theoretical conjectures about the scalar
glueball and show that the discovered BES state is a good glueball
candidate meriting more careful study. Using a QCD-based model we
calculate that the full glueball decay width is about 100 MeV which is
consistent with the measured 105 $\pm$ 20 MeV width for the
$f_0(1810)$.  We also show that the commonly assumed flavor blind
glueball decay treatment entails large corrections yielding a
measurable $\omega \phi$ branch but a suppressed $\omega \omega$
channel, again consistent with new data. Finally we demonstrate how
the rescattering mechanism, $f_0(1810) \rightarrow K{\bar K}
\rightarrow$ $\phi \omega$, facilitates observing $\phi \omega$
cleanly above the tail of the predominant $f_0(1710) \rightarrow
K{\bar K}$ spectrum, given sufficient precision.

\section{Phenomenological considerations}

\subsection{Contrasting BES with DM2 and Mark III data}

About two decades ago the DM2 collaboration observed at Orsay
$8.6\times 10^6$ $J/\psi$ decays and studied the $\omega\omega$
spectrum triggered by radiative decay (photon decays have a 1\% branch
due to $\alpha_{EM}/\alpha_s$ suppression). They reported
\cite{Bisello:1987ht} a branching ratio,
$B(J/\psi\to \gamma\omega\omega) = 1.4(0.2)(0.4) \times 10^{-3}$, much
larger than the BES $\gamma\omega\phi$ ratio, $B(J/\psi\to
\gamma\omega\phi) = 2.61(0.27)(0.65) \times 10^{-4}$, and observed a
strong pseudoscalar $\eta(1760)$ enhancement but concluded there was
no relevant structure near or above 1800 MeV. Of course the
$\omega\omega$ decay must correlate with about 3 $\rho\rho$ decays and
indeed Ref. \cite{markIIIprd33} observed a similar pseudoscalar signal
in $\rho \rho$ about 100 MeV wide, but threshold effects make it
difficult to compare these two experiments.  Related, there should
also be a correlation with the $\omega \phi$ channel
\cite{Cotanch:2004py,Cotanch:2005ja} (note DM2 was not designed to
detect $\omega \phi$) however this can only be observed for resonances
above the $\omega \phi$ 1802 MeV threshold (it would also be
kinematically suppressed unless significantly above threshold).

Most recently, the BES collaboration also reported \cite{bes2} an
analysis of $J/ \psi \rightarrow \gamma \omega \omega$ from the same
5.8 $\times 10^7 J/ \psi$ BESII detector data set.  They clearly
confirm the $\eta(1760)$ (mass 1744 MeV, width 244 MeV) and also
evidence for a $0^{++}$ structure which could correspond to the
$f_0(1710)$ and/or $f_0(1810)$.  However, it is difficult to determine
the mass and width of this scalar hadron due to the dominant
contributions from the $\eta(1760)$.

\begin{figure}
\psfig{figure=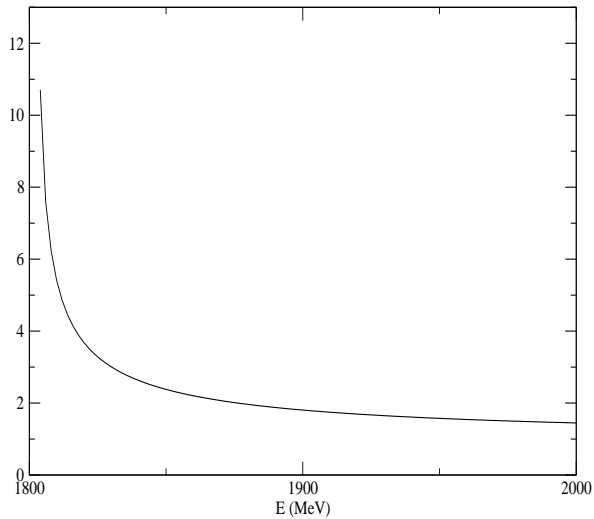,width=3.5in,height=3.7in,angle=-90}
\hspace{-1.9cm}
\caption{\label{phasespaceplot} Ratio of $\omega \omega$
to $\omega\phi$ for a Breit-Wigner folded with phase space.}
\end{figure}

\begin{figure}
\psfig{figure=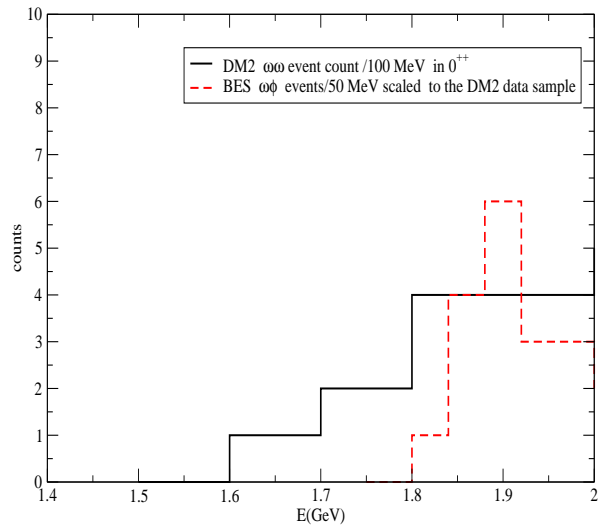,width=3.5in,height=3.7in,angle=-90}
\caption{\label{bestodm2} 
Dashed line: BES $\omega\phi$ events rescaled to the DM2 sample size.
Solid line: actual $0^{++} \omega\omega$ DM2 measurement.  A DM2
confirming $\omega\phi$ decay signal is not be possible due to their
small $J/\psi$ sample. }
\end{figure}

We submit the absence of the $f_0(1810)$ scalar resonance, independent
of its quark-glue structure, in the DM2 and BESII $\omega\omega$
spectra indicates the $f_0(1810)$ decay is not flavor blind.  This is
because the BESII $\omega \phi$ signal, combined with flavor
independent decay (omitting phase space effects), predicts an
$\omega\omega$ signature that would have also been observed by BESII
and DM2.  To quantify, we compute the ratio, $R$, of $\omega \omega$
to $\omega \phi$ phase space factors, $P(m)$, folded with a
Breit-Wigner (see Fig.~\ref{phasespaceplot}),
\be
R = \frac{\int_{\omega\omega~{\rm th}}^{2\ {\rm GeV}}dm 
\frac{P_{\omega\omega}(m)}{(m-M_{f_0})^2+\Gamma_{f_0}^2/4}}{\int_{\omega\phi~{\rm th}}^{2\ 
{\rm GeV}}dm  \frac{P_{\omega\phi}(m)}{(m-M_{f_0})^2+\Gamma_{f_0}^2/4}} \ .
\ee
Multiplying $R$ by the number of BES $f_0(1810)$ observed events (95)
and the ratio of DM2 to BES $J/\psi$ total decays ($8.6\times
10^6$/$5.8\times 10^7$) yields 58 $f_0 \rightarrow \omega\omega$
events DM2 would have reported, assuming equal reconstruction
efficiencies and a $f_0(1810)$ flavor blind decay.  However, as
detailed in Fig.  \ref{bestodm2}, DM2 only reported 4 $\omega\omega$
events in the $0^{++}$ channel, which undermines the $f_0(1810)$
flavor blind decay assumption.  Predictions for other possible $f_0$
mass assignments are summarized in Table
\ref{tableDM2}.
\begin{table} [b]
\begin{tabular}{|cc|cc|}
\hline
$M_{f_0}$ & $\Gamma$ & R & Predicted DM2 Events \\ \hline
 & 0.105 & 3.9 & \\
1.812 & 0.085 & 4.1 & 56 \\
 & 0.125 & 3.8 & \\
\hline
1.793 & 0.103 & 5.0 & 71 \\
1.838 &0.105 & 3.0 & 43 \\
\hline
\end{tabular}
\caption {$\omega\omega$ events DM2 would have observed in the $0^{++}$ channel assuming
flavor blind decay.  Only 4 events were recorded by DM2 indicating a
preference for open flavor decay. Units for $\Gamma$ and $M_{f_0}$ are
GeV.}
\label{tableDM2}
\end{table}
Also, rescaling the BES data sample size for appropriate comparison
(see dashed line in Fig. \ref{bestodm2}), yields at best only a few
(less than 4) $\omega\phi$ events that the DM2 collaboration would be
expected to observe.  It would appear that the number of events in the
DM2 $J/\psi$ sample is insufficient to determine whether a resonance
is or is not present in their $\omega\phi$ spectrum even if they were
looking for it.

The DM2 data is in overall agreement with the Mark III data 
at SLAC (see \cite{Dunwoodie:1997an} and references therein). The Mark III $K\bar {K}$ 
and $\pi\pi$ spectra  featured a prominent $f_0(1710)$ but 
no $f_0(1810)$ state. 
Somewhat perplexing, BES  reported a low statistics study of the 
$K^*\bar{K}^*$ spectrum in radiative 
$J/\psi$ decays \cite{Bai:1999mk}
with the $0^{++}$ channel  not significantly populated.
We also note that the more recent
294 $\gamma \omega \phi$ events cleanly
isolated by BES, although only part of the total produced, represent  
an extremely small branching fraction compared to, for example,  $B(J/\psi \to
\gamma K^*\bar{K}^*)=4(1)\times 10^{-3}$.

The $f_0(1810)$ decay profile is perplexing. While suppressed rates to
the $K^*\bar{K}^*$ channel can be understood (note conservation of
$J^P$ forbids decay to $K^* \bar{K}$ and $\rho \pi$) due to limited
phase space (see next section), what scalar hadron would decay leaving
a clear signal in $\omega\phi$ but apparently none in either of
$K\bar{K}$, $\pi\pi$ or, most significantly, $\rho \rho$ and
$\omega\omega$?  In the next section we explain the suppressed $\rho
\rho$ and $\omega\omega$ decays by arguing that open flavor
(strangeness) glueball decays are favored and that $K\bar{K}$
rescattering plays an important role in the $\omega\phi$ final state.

\subsection{Radiative charmonium decay and glueball formation}

Consider the radiative $J/\psi$ decay to a $C=+1$ charmonium (on or
off-shell) that subsequently decays. Having positive $C$ parity favors
decay via intermediate two gluon states and the resulting spectra
should therefore display resonances corresponding to the glueball
masses.
\begin{figure}
\psfig{figure=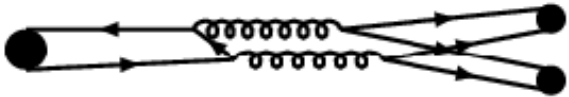,height=1in}
\caption{Depiction of a $C=+$ charmonium decay via gluonic and quark intermediate time steps 
(other orderings are possible).}
\label{Jdecay}
\end{figure}
A simple diagrammatic analysis (see Fig. \ref{Jdecay}) reveals that
open (explicit) flavor decays, that we call ``fall apart", dominate
over closed (hidden) flavor decays that require color exchange.  Here
the time axis is horizontal and the (on or off-shell) decay sequence
is: a charmed hybrid, glueball, light hybrid and finally a tetraquark
system. This yields a preference for open flavor mesons
(e.g. pseudoscalar or vector kaons) over closed (hidden) flavor
$\omega\phi$ that requires final state rescattering.  Table
\ref{pospardecays} further supports this point, listing established $C
= + 1$ charmonia decays \cite{Eidelman:2004wy} to predominantly open
flavor meson states. The data is best interpreted by assuming a
``fall-apart" decay mechanism with open flavor dominating over the
closed strangeness decay that requires color exchange (rescattering).
Where the $\phi\phi$ decay is unknown we enter the four kaon
decay. Likewise where the $K^* \bar{K}^*$ ratio is unknown we listed
the branching fraction to two charged pions and two charged
kaons. This enables estimating upper bounds for the two-body decays.
\begin{table} [b]
\begin{tabular}{|c|cccc|}
\hline
Channel & $0^{-+}\ (\eta_c)$ &  $0^{++} \ (\chi_c^0)$
&  $1^{++}\  (\chi_c^1)$ &  $2^{++} \ (\chi_c^2$) \\ \hline
$K^*\bar{K}^*$   & 8.5(3) & & & \\
$K^+ K^- \pi^+ \pi^-$ &  & 21(5) & 5(1) & 12(4) \\
$K^+\bar{K}^{*0}\pi^- \ +$ cc & & 12(4) & 32(21) & 48(28) \\
\hline
$\omega \omega$ & $<$3 & & & \\
$\phi \phi$ & 2.6(9)& 1.0(4)(4) & & 2.4(6)(7)  \\
$K^+ K^+K^- K^-$ &  & & 0.42(15)(12) & 1.7(3)(3)\\
\hline
\end{tabular}
\caption{Selected $C = + 1$ charmonium branching fractions with explicit and hidden
strangeness. All numbers should be multiplied by $10^{-3}$.}
\label{pospardecays}
\end{table}

\section{Glueball decay widths}
\subsection{Existing estimates }
In previous work we and others have published estimates for glueball
widths which we now summarize before presenting new computations.  As
detailed in Refs.~\cite{Cotanch:2004py,Cotanch:2005ja} the width for
the decay of a scalar glueball $G$ to two vector mesons, $G
\rightarrow V V'$, is
\be
\Gamma_{G \rightarrow V V'} = \frac{g_{GVV'}^2}{4\pi}\frac{k^3}{M^2}\; ,
\ee
where $M = 1$ GeV is a fixed reference mass, $g_{GVV'}$ is the $G VV'$
coupling constant and $k$ is the $cm$ momentum for the decay vector
mesons, given by $ k = (M_G/2) [(1 + x -x^\prime)^2 -4x]^{1/2},$ with
$x = (M_V/M_G)^2$, $x^\prime = (M_{V^\prime}/M_G)^2$ and $M_G$ the
scalar glueball mass (now tested against 1812 MeV).  Using Vector
Meson Dominance (VMD), Ref. \cite{Cotanch:2005ja} obtained $g_{GVV'} =
4.65$, which gives a small partial width of 1.43 MeV for the $\omega
\phi$ decay reflecting the near threshold suppressed phase space.  An
independent work \cite{Burakovsky:1998zg} reports a similar value for
the coupling, $g_{GVV'} = 4.23$.

If we assign the larger glueball mass $M_G = 1812 +19 = 1831$ MeV,
using the BES quoted upper statistical error, the $\omega \phi$ width
increases to $7.16$ MeV.  For the maximum possible mass (including the
quoted 18 MeV systematical error) of 1849 MeV the width further
increases to 15 MeV.  Table \ref{widthstable} lists predictions for
other two-body decays along with parameter sensitivity.  Relying on
these results alone would suggest the glueball width is much broader
than the $f_0(1810)$ BES candidate and its decay branching fraction to
$\phi\omega$ is insignificant.

\begin{figure}
\psfig{figure=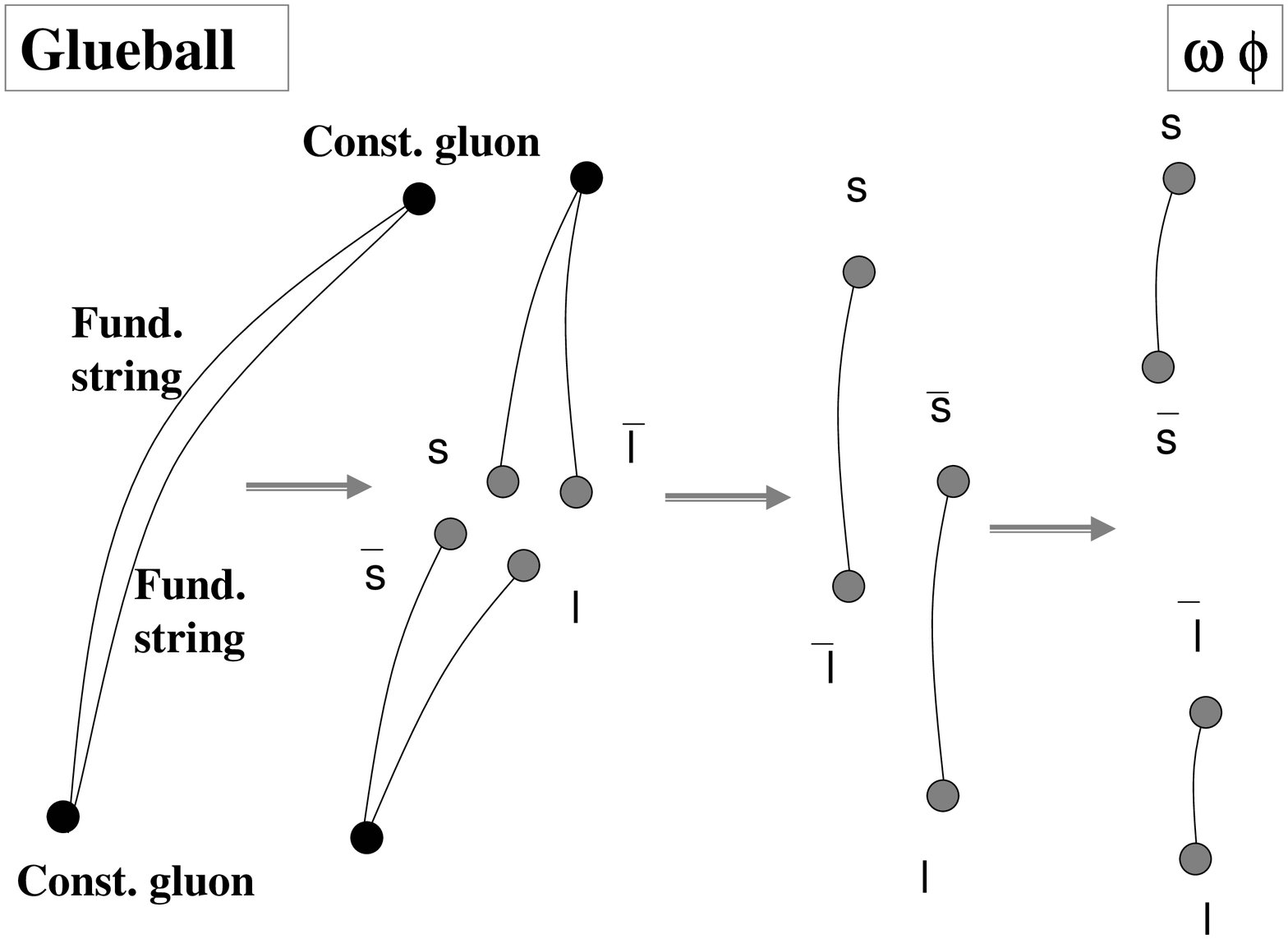,height=2.7in}
\caption{Illustrating glueball decay with final state rescattering of 
the produced kaons to yield $\omega\phi$.\label{breakinstring}}
\end{figure}

\begin{table} [b]
\begin{tabular}{|cc|cccc|c|}
\hline
$M_{f_0}$ & $g_{GVV^\prime}$ & $\Gamma_{\omega\phi}$ & 
$\Gamma_{\rho\rho}$ & $\Gamma_{\omega\omega}$ & $\Gamma_{K^*K^*}$ & $\Gamma_{\rm tot}$ \\
\hline
1700 & 4.65 & {\rm N/P} & 72.1 & 62.8 & N/P  & 135 \\
1812 &      & 1.43      & 176  & 164  & 4.10 & 346 \\
1831 &      & 7.16      & 198  & 184  & 11.3 & 401 \\
\hline
1700 & 4.23 & {\rm N/P} & 59.7 & 52.0 & N/P  & 112 \\
1812 &      & 1.18      & 146  & 135  & 3.39 & 286 \\
1831 &      & 5.92      & 164  & 153  & 9.38 & 331 \\
\hline
\end{tabular}
\caption{Partial widths (in MeV) for
various values of the glueball mass and coupling.  N/P indicates
insufficient phase space.  The final column is the sum of the other
columns and represents a lower bound for the total width.}
\label{widthstable}
\end{table}
Another glueball approach~\cite{Abreu:2005uw}, based upon a string
model, predicts a different, smaller total width, $\Gamma_{G}=140 $
MeV. The decay mechanism (with inelastic rescattering of the kaons in
the final state) in this constituent model is illustrated in Fig.
\ref{breakinstring}.
Their total width is much narrower than expected from VMD but more
comparable to the width of the BES candidate. In view of this
disagreement between simple string and VMD estimates we have performed
a new, first-principles calculation of the glueball width which is
described in the next section.

\subsection{Ab-initio glueball width computation}

A more fundamental, QCD-based calculation for the total width can be
obtained using many body theory~\cite{Llanes-Estrada:2005jf}.  In this
relativistic, field theoretical approach an effective Coulomb gauge
QCD Hamiltonian is approximately diagonalized using the BCS and TDA
many body treatments for the vacuum (gap equation with dressed gluons)
and hadron states, respectively.  Results are briefly described with
further details relegated to Appendix \ref{widthcalc}.

The glueball is represented by the lightest Fock state consisting of
two constituent BCS transverse gluons which decay to two quark pairs
that subsequently hadronize.  The decay matrix element is
\be \label{Matelement}
\mathcal{M} = \la G \ar \frac{1}{2} \int d {\bf x} d {\bf y} H_{qg}({\bf x}) 
G({\bf x},{\bf y}) H_{qg}({\bf y}) \ar qq\bar{q}\bar{q} \ra\ ,
\ee
where the quark-gluon Hamiltonian field interaction is specified in
the appendix and $G({\bf x},{\bf y})$ is the propagator for
intermediate scalar hybrid meson states.  The scalar glueball state
involving BCS quasi-gluon creation operators
$\mbox{\boldmath$\alpha$\unboldmath}^{\da a}$, color index $a $,
operating on the BCS vacuum $|\Omega \ra$ is
\be \label{glueballdef}
\ar G \ra = \int \frac{d  \bs k}{(2\pi)^3} 
\frac{\phi(\bs k)}{\sqrt{4\pi}} 
\frac{\sqrt{M_G} }{4}
\mbox{\boldmath$\alpha$\unboldmath}^{\da a}(\bs k)\cdot  \mbox{\boldmath$\alpha$\unboldmath}^{\da a}(-\bs k) \ar \Omega \ra \ ,
\ee
with wavefunction normalization
\be
\int  \frac{d \boldsymbol k}{(2\pi)^3} \frac{\ar \phi(\bs k)\ar^2}{4\pi} =1\ .
\ee
The unit normalized quark state is
\be
\ar {\bf q} \lambda\ra = \sqrt{2E} \sum^3_{ {\cal C} = 1} B^{\da}_{\lambda \cal{C}}({\bf q})
\hat{ \epsilon}_{\cal C} \ar \Omega \ra \ ,
\ee
for dressed quark creation operator $B^{\da}_{\lambda \cal{C}}({\bf
q})$ and color vector $\hat{ \epsilon}_{\cal C}$.  Finally, the width
is given by
\be
d\Gamma= \frac{1}{2M_G} \ar \mathcal{M} \ar^2 d\Phi_4 \ .
\ee
where the four-body phase space for the final quarks is
\be \label{intphasespace}
d{\Phi_4}=(2\pi)\delta^4(M_G -\sum_{i=1}^4 {E_i}) \left(
\prod_{i=1}^3 \frac{d \bs q_i}{(2\pi)^32E_i}
\right) \frac{1}{2E_4} \ .
\ee
Consult Appendix \ref{widthcalc} for the remaining, technical details
of this large-scale, multidimensional integral calculation.  However
using dimensional analysis immediately reveals that the total width is
of order 100 MeV. Numerical predictions are listed in Table
\ref{inclusivewidth}. The first column is our reference calculation
and lists the widths for a glueball with mass 1812 MeV and a flavor
independent quark-gluon coupling vertex. The second column is for a
flavor dependent, and stronger, $ssg$ vertex inspired by a Landau
gauge study in which resummed, leading $N_c$ radiative corrections
were more suppressed for light quarks.  The dependence on flavor
factors follows directly from Eq. (\ref{finalforthewidth}).  The third
column illustrates sensitivity to the glueball wavefunction.  The
calculation is the same as the first column except the TDA
wavefunction is taken from Ref. \cite{Llanes-Estrada:2000jw}, where a
slightly lighter scalar glueball mass of about 1725 MeV was calculated
(however we maintain the BES 1812 MeV kinematics/phase space).
Finally the fourth column illustrates sensitivity to the momentum
cut-off used in the calculation and represents probability flux
leaking to other channels.  Wavefunction components leading to total
momentum/energy above the mass of the decaying glueball are virtual
and suppress the width. Eliminating them by artificially reducing the
cutoff in the glueball wavefunction to $M_G/2$ increases the width by
about a factor of 2 which is also the upper bound to cut-off
sensitivity.
\onecolumngrid

\begin{table} [b]
\begin{tabular}{|c|c|c|c|c|}
\hline
Widths (MeV) & Flavor independent $qqg$ vertex & $ssg\simeq 
\frac{10}{7}uug\ (ddg)$  & $G$ wf. from 
\cite{Llanes-Estrada:2000jw}, $\Lambda=8$ GeV & $G$ wf. from
\cite{Llanes-Estrada:2000jw}, $\Lambda=0.9$ GeV
\\ \hline
$\Gamma_{\rm tot}$ &       100 & 175 & 50 & 90 
\\ 
$\Gamma_{\rm light-light}$ &     50 & 50 & 25 & 45
\\
$\Gamma_{\rm light-strange}$ &   30 & 65 & 15 & 30
\\
$\Gamma_{\rm strange-strange}$ &  15 & 60 & 5 &  10
\\
\hline
\end{tabular}
\caption{
Total and partial widths for a scalar glueball with mass 1812 MeV.
Light refers to a light $u/d$ quark-antiquark pair and strange denotes
a $s \bar{s}$ pair. }\label{inclusivewidth}
\end{table}

\newpage
\twocolumngrid
Note that in contrast to the above phenomenological models, the
calculated total widths from the more fundamental theory are narrower,
of order 100 MeV, and consistent with the BES measurement.  Indeed,
our result also affirms an argument for narrow glueball widths
published \cite{Carlson:1981ve} some time ago based on the OZI rule
and originally applied to the oddball (three gluon $1^{--}$ glueball,
see \cite{Llanes-Estrada:2005jf}). The assertion was that charmonia
decay dominantly via a glueball/oddball intermediate state, which in
turn selects light hadron decay channels, so that the actual width of
the glueball is about the geometric mean of the width of OZI-allowed
and OZI-suppressed decays, of order a few tens of MeV.

Concluding this subsection, our best estimate for the total glueball
width is about 100 MeV.

\subsection{Width ratios including final state rescattering}

We have also evaluated final state effects and present further details
in Appendix \ref{RGM}.  Here we qualitatively comment and focus on a
simple rearrangement potential between the $K \bar K$ and $\phi
\omega$ channels.  This flavor exchange, contact potential couples
different channels and illustrates how the $\omega\phi$ signal can
arise from other channels by final state rescattering.  Table
\ref{vretable} lists the ratio of channel 2 to 1 partial widths for
different potential strengths.
\begin{table} [b]
\begin{tabular}{|c|cccc|}
\hline
$1\to 2$ & $\ar {\rm Spin}\ar^2$ & $\ar {\rm Flavor}\ar^2$ & $\ar v_{re}\ar$ & $\Gamma_2/\Gamma_1$ \\ \hline $K^*\bar {K}^*\to \phi\omega$ & 
     25/4  &  2   & 100 MeV  & 0.044  \\
$K \bar{K}\to \phi\omega$ & 
     3       &  2   & 69 MeV  & 0.006  \\
$\rho\rho\to \omega\omega$ & 
     25/4  &9/4 & 106 MeV   & 0.020  \\
$\pi\pi\to \omega\omega$ & 
     3       &9/4  & 73 MeV  & 0.011 \\ 
\hline
\end{tabular}
\caption{\label{vretable} Rearrangement potential factors and ratio of widths for different channels using $V_{SS} \simeq 200$ MeV. See Appendix \ref{RGM} for details.}
\end{table} 
%
The widths are calculated using second order perturbation theory which
should be reasonable as long as the width ratio remains below 1. Also
note that $K^*\bar{K}^*$ and $\rho\rho$ rescattering effects are
suppressed by their large widths (50 and 150 MeV, respectively) which
will broaden any signal and are thus not relevant to the narrow BES
state. More promising is the $K\bar{K}\to \omega\phi $ rescattering
process which is somewhat smaller but still sizeable.

Other factors explain why the $\pi\pi\to\omega\omega$ process is not
important. Whereas the two processes
\ba\nonumber
G\to K \bar K \to \omega\phi\\ \nonumber
G\to \pi\pi \to \omega \omega
\ea
have similar rescattering strengths, there is a factor of about 2
suppression with respect to the $\omega\phi$ due to the stronger
strange quark coupling, and an additional factor of about 4 from
wavefunction overlap suppression due to the very different scales
involved. This reduces the relative rescattering contribution to
$\omega\omega$ by almost an order of magnitude.  An $f_0(1810)$ signal
in this channel would not be observed by DM2, and at best marginally
with the BES statistics.

As for the current absence of a BES $f_0(1810)$ signal in $K\bar{K}$,
we submit a more extensive measurement will observe this decay.  This
should include a careful examination for any enhancement in the tail
of the established $f_0(1710) \rightarrow K\bar{K}$ decay.  Related,
and as first pointed out in
Refs. \cite{Cotanch:2004py,Cotanch:2005ja}, the $\omega \phi
\rightarrow K \bar{K} 3 \pi$ is a distinctive, novel glueball
signature easily detected.  It may be that the other decay channels
were more difficult to observe due to pion background effects
(e.g. $\rho \rho \rightarrow 4 \pi$ and $\omega \omega \rightarrow 6
\pi$ and even $\eta \eta, \eta \eta' \rightarrow $ multiple $\pi$).


\section{Alternative  {\boldmath $ \lowercase{f}_0(1810)$} scenarios}
\subsection{Threshold cusp}

We first examine the possibility that a threshold cusp
\cite{Bugg:2004rk} explains the structure in the $\omega\phi$ 
spectrum. This kinematical enhancement occurs when a two-body system
inelastically couples strongly to another open channel near threshold.
Even at 1.8 GeV this condition is possible, and this would produce a
low momentum scattering amplitude having form $A+B/k$ with $k$ the
$\omega \phi$ center of mass momentum. However, multiplying the BES
data by kinematical factors appropriate to each bin yields a resonance
that is well separated from threshold which seems to rule out this
option.

Furthermore, the DM2 $\omega \omega$ and $\rho\rho$ data should have a
similar cusp but there are none. Rather, these data reveal a prominent
peak, the $\eta(1760)$, 200 MeV above threshold, and clearly
monotonically fall towards lower energies.

\subsection{Conventional or hybrid meson}

Even though there have been many scalar meson studies, their structure
is still not completely understood. In the absence of mixing (claimed
to be significant in most analyses), quark model $f_0$ states have two
isoscalar flavor combinations, $s \bar{s}$ and $n \bar n
=(u\bar{u}+d\bar{d})/\sqrt{2}$, with $ ^{2S+1}L_J= \, ^{3}P_0$. Their
ground states are slightly above 1 GeV \cite{Llanes-Estrada:2001kr},
and for this argument we use 1.1 and 1.4 GeV for the light and strange
quark combinations, respectively.  Adding 500-600 MeV for the required
radial excitation (e.g. $\phi(1020)$ and its radial excitation
$\phi(1680)$) yields 1.6-1.7 GeV for the light, and 2 GeV for the
strange combination.  The light quark combination is marginally too
low while the $s\bar{s}$ radial excitation state is too high to
explain a resonance at 1.8 GeV.  Moreover, one would expect the latter
to have a sizeable $K\bar{K}$ branching fraction, but this is not
visible in the Mark III data \cite{Dunwoodie:1997an} where the
$f_0(1710)$ is dominant. Mixing the $f_0(1710)$ with a $n \bar n$
radial excitation may perhaps explain the BES peak. If so its decay to
$f_0(1370)\pi\pi$ might be visible, however there is no simple
mechanism explaining why this state should appear in $J/\psi$ decays.
Although this assignment cannot be rejected, these arguments make us
suspect.

We next examine hybrid mesons.  In many body theory
\cite{Llanes-Estrada:2000hj}, hybrid mesons with the minimal Fock
space assignment $q\bar{q}g$ in an s-wave yield a triplet
$(0,1,2)^{++}$ and a pseudovector $1^{+-}$. However, for typical
values of the string tension, $\sqrt{\sigma}=367$ MeV, their masses
are near but above 2 GeV. Similar, though a bit lighter, results are
obtained in the flux tube model and lattice gauge theory, so one
cannot discard a hybrid state as low as 1.8 GeV. However qualitatively
comparing hybrid and glueball total width calculations, the hybrid
width will be much broader since there is only one gluon-quark vertex
interaction, instead of two, yielding one less factor of $\alpha_s^2$
suppression.  Therefore a broader state than the BES result is
expected.

\subsection{The {\boldmath $ \lowercase{f}_0(1710)$} tail}

The mass and width of the $f_0(1710)$, another glueball candidate, are
poorly determined and the PDG values are $M=1714(5)$ MeV,
$\Gamma=140(10)$ MeV. However, consistent with recent BES data, these
values could be as high as $M=1740$ MeV, $\Gamma=166$ MeV. In this
case, an overlap with the trailing edge of the $f_0(1710)$
Breit-Wigner distribution could produce the observed $\omega\phi$
signal. However this possibility appears unlikely considering the
near-threshold behavior of the BES $\omega\phi$ spectrum. As in the
cusp hypothesis, the current data seems to indicate that the resonance
is separated from the threshold and therefore cannot stem from the
$f_0(1710)$. Higher sample count studies would be very useful.

\subsection{Four quark states}

Tetraquark systems, another actively investigated area, also appear
naturally as an intermediate step in a $J/\psi$ decay chain.  However,
as in the hybrid case, one expects a four quark state to decay with a
broad width generating a background, not a sharp signal, for radiative
$J/\psi$ decay.  Related, a realistic tetraquark width prediction also
requires including $K \bar{K}$ rescattering effects since the $\omega
\phi$ attraction is not as strong as in $K \bar{K}$ where annihilation
diagrams provide attractive forces.  The quark rearrangement coupling
between the $K^* \bar{K}^*$ and $\omega \phi$ channels also provides
attraction (see Appendix B).  This follows from the Resonating Group
Method (RGM) \cite{sccm, Bicudo:2001jq, Bicudo:2003rw} which predicts
an increased attraction between mesons when each has a quark and
anti-quark of the same flavor.  Hence if the BES state is not a
glueball, the RGM coupled channels will play an important role in
elucidating its structure and applications of our model to this system
are in progress.
 
\section{Summary and conclusions}

In this work we have examined and compared independent $J/\psi$ decay
data sets in the 1800 MeV mass region. Based on the data reported by
the BES collaboration, we believe that the newly found $f_0(1810)$ is
a promising glueball candidate or a state with a large glueball
component. Significantly, its mass and quantum numbers are in
agreement with previous theoretical expectations and its somewhat
surprising narrow width of order 100 MeV is consistent with new
fundamental calculations. We have addressed the perplexing issue of
its selective decay to the $\omega \phi$ channel and discussed why it
was not observed in both the BESII and the smaller sample DM2
measurements in the $\omega \omega$ channel.  Also the $\pi\pi$ and
$KK$ channels have been examined by MARK III and only the broad
$f_0(1710)$ structure is apparent. However, the binning of these data
is somewhat coarse and further structure cannot be ruled out. The DM2
data \cite{Augustin:1987fa} shows a falling slope probably due to the
$f_0(1710)$ tail. Higher precision studies are clearly needed.

We also noted that the rescattering process, $J/\psi\to \gamma G\to
\gamma K \bar{K} \to \gamma \omega\phi$, may be producing the BES signal.
In view of the important theoretical implications of a glueball state,
we submit that a precision study of the $K\bar{K}$ spectrum is crucial
and may reveal a significant enhancement around 1812 MeV. Should this
test fail, we would be facing a new theoretical puzzle.

\begin{acknowledgments}

F. J. Llanes-Estrada appreciates inspiring discussions on $J/\psi$
decays with W. S. Hou and also on glueball widths with A. Goldhaber.
Work supported in part by Spanish grants FPA 2004-02602, 2005-02327,
PR27/05-13955-BSCH (FJL), U. S. DOE Grant No. DE-FG02-97ER41048 (SRC),
Research Corporation (DGR) and the Ohio Supercomputer Center (DGR).
\end{acknowledgments}

\newpage
\appendix
\section{Computation of the inclusive decay width}
\label{widthcalc}
Here we present the many body effective QCD Hamiltonian calculation
for the glueball decay to four quarks that subsequently hadronize.  In
the Coulomb gauge the effective quark Hamiltonian contains an
instantaneous interaction, mediated by the infrared enhanced Coulomb
potential, and a transverse gluon exchange interaction that is
infrared suppressed via the generation of a mass gap
\cite{Szczepaniak:2003ve}.  First the instantaneous interaction is
diagonalized to obtain the glueball bound state wavefunction. Then the
triple quark-gluon coupling interaction
\be
H_{qg} = g \int d \bs x \Psi^\da T^a \mb \Psi \cdot \bs  A^a
\ee
is treated perturbatively to compute the decay amplitude.  Omitting
the momentum conserving delta functions arising from the commutators,
the integrand, $I$ of the matrix element in Eq.  (\ref{Matelement})
reduces to
\ba \label{matelementfull}
I(G \to q_1 q_2 q_3 q_4) = {\rm CT}\ 
\sqrt{2} \frac{g^2(k)}{2\omega(k)}\times 
 \; \; \;\; \; \; \; \; \;\; \; \; \; \; \; \\ \nonumber
\sqrt{2M2E_12E_22E_32E_4} \frac{\phi(k)}{\sqrt{4\pi}} 
G S_{{\cal C}_1{\cal C}_2 {\cal C}_3 {\cal C}_4}({\bf  \hat{q}_1 \hat{q}_2\hat{q}_3\hat{q}_4}) \ .
\ea
For total (inclusive) decay the color tensor  is 
\be
{\rm CT}= \frac{\delta_{ab}}{\sqrt{8}} T^a_{{\cal C}_1 {\cal C}_2} T^b_{{\cal C}_3 {\cal C}_4} \, .
\ee
which when squared and summed over the quark color indices ${\cal
C}_i$ yields the squared color factor ${\rm CF}^2 = 1/4$. The above
result is for only one specific flavor and below we include the
modification for application to three light flavors (u, d, s). The
$\sqrt{2}$ factor is a result of gluon exchange symmetry and the
glueball normalization in Eq. (\ref{glueballdef}). The gluon
self-energy, $\omega(k)$, follows from the intermediate gluon
propagators in Fig. \ref{Jdecay} and is the solution of a mass gap
equation that is well approximated by $\sqrt{m_g^2+k^2}$ (used here).
The tensor $S$ depends on the spinors in the Fourier expansion of the
quark field $\Psi$ and the Dirac \mb matrices coupled to the gluon
spins.  These spinors are usually expressed in terms of a BCS angle,
whose relation to the running mass (from the quark gap equation) is
$\sin\phi (q)= s_q= m(q)/\sqrt{m(q)^2+q^2}$ (here we fix $m(q)=m$).
Squaring the matrix element and summing over spins in the final state,
we find, in terms of unit momentum vectors
\ba
\sum_{{\cal C}_1 {\cal C}_2 {\cal C}_3 {\cal C}_4} \ar S \ar^2 = (1+s_1s_2)(1+s_3s_4) \qquad\qquad \\
+ (1+s_1s_2)c_3c_4{ \bf \hat{k}\cd\hat{q}_3  \hat{k}\cd\hat{q}_4}
\nonumber
+(1+s_3s_4)c_1c_2{ \bf \hat{k}\cd\hat{q}_1  \hat{k}\cd\hat{q}_2 }
\\ \nonumber
+c_1c_2c_3c_4 \Bigl  [ {\bf  \hat{q}_1\cd\hat{q}_3  \hat{q}_2\cd\hat{q}_4
+\hat{q}_2\cd\hat{q}_3  \hat{q}_1\cd\hat{q}_4 -
\hat{q}_1\cd\hat{q}_2  \hat{q}_3\cd\hat{q}_4 }  \\ \nonumber
- {\bf\hat{k}\cd\hat{q}_2 \hat{k}\cd\hat{q}_4 \hat{q}_1\cd\hat{q}_3
-\hat{k}\cd\hat{q}_1 \hat{k}\cd\hat{q}_4 \hat{q}_2\cd\hat{q}_3
+\hat{k}\cd\hat{q}_1 \hat{k}\cd\hat{q}_2 \hat{q}_3\cd\hat{q}_4 }\\ 
\nonumber
+ {\bf \hat{k}\cd\hat{q}_3 \hat{k}\cd\hat{q}_4 \hat{q}_1\cd\hat{q}_2 
-\hat{k}\cd\hat{q}_2 \hat{k}\cd\hat{q}_3 \hat{q}_1\cd\hat{q}_4
-\hat{k}\cd\hat{q}_1 \hat{k}\cd\hat{q}_3 \hat{q}_2\cd\hat{q}_4 }\\ 
\nonumber
\left.
+2 {\bf \hat{k}\cd\hat{q}_1\hat{k}\cd\hat{q}_2 \hat{k}\cd\hat{q}_3 
\hat{k}\cd\hat{q}_4 }
\right ]  \ .
\ea
Since all possible relative angles in the final state appear, $I$ is
only invariant under rigid rotations of the five vectors. With this,
the squared decay matrix element summed over the final state color,
spin and flavor indices reads
\ba
\sum \ar I \ar^2 &=& {\rm FF}^2 R \nonumber \\
R&= &{\rm CF}^2 
\frac{2}{4\omega(k)^2} 2M 2E_1 2E_2 2E_3 2E_4 \nonumber \\
&& \times\frac{\ar \phi(k)\ar^2}{4\pi} (4\pi \alpha_s(k))^2 \ar G S_{1234}\ar^2 \ .
\ea
Symmetry considerations apply if both emitted quark-antiquark pairs
are indistinguishable and we absorb this into the flavor factor ${FF}
^2$ below.

Let us now examine the phase space integrals. Momentum conservation
for a decaying glueball at rest requires
\be
{\bf q}_1 + {\bf q}_2 + {\bf q}_3 +{\bf q}_4 = { 0} \ ,
\ee
yielding for the gluon momentum 
\be
{\bf k} = {\bf q}_1 + {\bf q}_2 = -{\bf q}_3 - {\bf q}_4 \ .
\ee
There are nine integration variables in Eq. (\ref{intphasespace}). We
can arbitrarily fix $\bf{k}$ along the third axis and obtain a $4\pi$
factor for global rotations. Further we can integrate one azimuthal
angle (say $\phi_1$) around this fixed axis, an operation preserving
all relative angles. The modulus $k= \ar \bf{k} \ar$ remains an
independent variable. The others can be chosen as $q_1=\ar
\bf{q}_1\ar$, $\cos \theta_{q_1k}$ (which automatically fixes
$\bf{q}_1$ and $\bf{q}_2 $), and the three spherical coordinates of
$\bf{q}_3 $, $q_3=\ar \bf{q}_3\ar$, $\cos \theta_{q_3k}$ and
$\phi_3$. Only five of these six are independent since we have not
utilized the energy conservation relation. This imposes cumbersome
restrictions on the angular variables, so it is convenient and
customary to introduce an auxiliary variable, $\cal E$, representing
the energy of the second pair, by means of
\be
\delta(M_G -\sum E_i)=  \int \delta(M_G -  {\cal E}- E_1 - E_2)
\delta({\cal E}-E_3-E_4)d{\cal E} \  . \nonumber
\ee
The two $\delta$ functions can be used to integrate over the two polar
angles $\cos\theta_{q_1k}$ and $\cos \theta_{q_3k}$, leaving only the
${\cal E}$ integration with integration limits fixed by the
requirement that the cosine values remain in the interval
$(-1,1)$. This is easily implemented in our 5 dimensional Monte Carlo
computation by rejecting points exceeding this bound.  The resulting
polar cosines are
\ba
\cos \theta_{q_1k}&=&\frac{m_2^2+q_1^2+k^2-(M_G - {\cal E}+E_1)^2}{2kq_1}
\\ 
\cos \theta_{q_3k}&=&\frac{({\cal E}-E_3)^2-(m_4^2+q_3^2+k^2)}{2kq_3} \ .
\ea
Note that the change of variable from energy to angle in each of the
$\delta$ functions adds an extra factor
\ba
\delta(E_0-E)&=&\delta(E_0 -\sqrt{m^2+k^2+q^2+2kq\cos \theta})
\nonumber \\
&=&\frac{E_0}{kq} \delta(\cos \theta_0 -\cos \theta) \ .
\ea

There are then four remaining integration variables $k$, $q_1$, $q_3$,
$\phi_3$, for a total of five integrals that are performed
numerically.  A representation for the coupling $\alpha_s$ in the
infrared is needed and we use \cite{Brodsky:1982nx}
\be
\label{alphas}
\alpha_s(k)=\frac{4\pi}{9\log((k^2+M_0^2)/\Lambda^2)}
\ee
with $\Lambda\simeq 0.2-0.21 $ GeV and $M_0 \simeq 1-1.1$ GeV.  The
final ingredient is the propagator $G(\bs x, \bs y)$ for the
intermediate hybrid meson cut in Fig. \ref{Jdecay}, necessary for a
second order calculation. Its exact energy eigenfunction expansion is
\be
G =\sum_{\rm h = hybrid} \ar h \ra\frac{1}{M_G - E_h -i\epsilon} \la h   \ar\ .
\ee
The spectrum of hybrid mesons has been studied with this many body
method in \cite{Llanes-Estrada:2000hj} where, for string tension
$\sqrt{\sigma}=367$ MeV, the ground state scalar hybrid meson has mass
$2100$ MeV. Excitations thereof appear with spacings similar to those
in ordinary meson quark models. We use $G=1/(M_G-M_h)\simeq $1/(300
MeV) and dressed quark masses $m_u=m_d=100$ MeV, $m_s=200$ MeV,
consistent with prior work using the same approach and parameters.
These values are typical of the masses calculated in Ref.
\cite{Llanes-Estrada:2004wr}, but somewhat low compared to quark model
phenomenology since in field theory approaches a sizeable fraction of
the hadron mass originates from the self-energy contribution in the
bound state problem and not in the mass gap equation. These values
also yield a realistic conventional hadron spectrum. Similarly we have
$m_g=\omega(0) = $ 650 MeV, where $\omega(k)$ is the solution of the
gluon gap equation of pure gluodynamics \cite{Szczepaniak:1995cw} .

Finally let us examine the flavor factors.  For an inclusive decay we
can separate the sum over the final states
\be
\sum_{\rm flavor} \ar I \ar^2= \sum_{\rm diff}  \ar 
I \ar^2 + \sum_{\rm same}  \ar I \ar^2  = FF^2 R \; ,
\ee
where
\ba
FF \propto \la \Omega \ar (u\bar{u} + d\bar{d} + s\bar{s}) 
\sum_{q'\bar{q}'} \ar q'\bar{q}' \ra \\ \nonumber
\la q'\bar{q}' \ar  (u\bar{u} + d\bar{d} + s\bar{s})  \ar q\bar{q} q\bar
{q} \ra\; .
\ea
We have
\be
FF^2 = 4\cdot 3 + \frac{1}{2^2}\cdot 16\cdot 3 = 24\; ,
\ee
where the first term accounts for the case where the outgoing quark
pairs have different flavors, and the second for the case where the
outgoing flavors are the same.  The 3 in each term reflects the number
of distinct choices for three flavors $(u, d, s),$ and the $1/2^2$
corrects for over-counting in the sum over final states with two pairs
of identical particles.  If we separate by flavor channel the
corresponding factors are 12 for light-light, 8 for light-strange and
4 for strange-strange.  With this we obtain the complete expression
for the glueball width (see Table~\ref{inclusivewidth} for numerical
results)
\begin{widetext}
\ba \label{finalforthewidth}
\Gamma= \frac{(2\pi)(4\pi)}{(2\pi)^9} \int_{0}^{M_G} 2\pi d{\cal E}
\int_0^{2\pi}d\phi_3\int_0^{M_G/2} k^2dk \int_0^{M_G/2} q_1^2dq_1 
\int_0^{M_G/2} q_3^2 d q_3\qquad\qquad \\ \nonumber
\times\frac{E_2}{kq_3} \frac{E_4}{kq_1} \frac{1}{(M_G-M_h)^2} {\rm CF}^2
{\rm FF}^2 \frac{2}{4\omega_k^2} \frac{\ar \phi (k)\ar^2}{4\pi}
\ar S\ar^2 (4\pi \alpha_s(k))^2 \Theta(\cos^2\theta_{q_1k}-1) 
\Theta(\cos^2\theta_{q_3k}-1) \, .
\ea
\end{widetext}

\section{Resonating Group Method and decay channel recoupling}
\label{RGM}

In this appendix we theoretically treat the sequential decay of a
glueball $G$ to a meson pair followed by rearrangement. To be
specific, we assume that $G$ first decays to $K^* \bar K^*$ and then
to $\phi \omega$ as depicted in Fig. \ref{breakinstring}.

\subsection {Coupled channels}

We approximately solve the equation of motion, $H\Phi = E\Phi$, using
the resonating group/coupled channels formalism
\cite{sccm,Bicudo:2001jq,Bicudo:2003rw} for this three channel problem 
\begin{widetext}
\begin{eqnarray}
&&
\left( 
\begin{array}{ccc}  
 H_G-E-i \epsilon & V_{sb} & 0 \\
 V_{sb}^* & H_{K^* \bar K^*} -E-i \epsilon &  V_{re} \\
0 & V_{re}^* & H_{\phi \omega} -E-i \epsilon 
\end{array}
\right) 
\left( 
\begin{array}{c}  
 \Phi_G  \\
 \Phi_{K^* \bar K^*} \\
 \Phi_{\phi \omega}  
\end{array}
\right) 
=0
\ ,
\label{eq:coupled}
\end{eqnarray}
\end{widetext}
where $V_{sb}$ is the string breaking decay coupling between the
glueball and the open flavor channel, and $V_{re}$ is the
rearrangement potential coupling the latter to the ${\phi \omega}$
channel.  We now extract the glueball width $\Gamma$ from the
imaginary part of the resulting glueball energy/mass.  Since
$H_G\simeq M_G = 1812$ MeV and from our computation of the total
glueball width, it follows that $V_{sb}$ is at most of order 100 MeV.
We can therefore diagonalize using perturbation theory to the leading
order in $V_{sb}$,
\begin{eqnarray}
&& \biggl[ H_G-E- V_{sb} \biggl(  { 1 \over H_{K^* \bar K^*} -E-i 
\epsilon } 
\nonumber \\
&&+ { 1 \over H_{K^* \bar K^*} -E-i \epsilon } 
 V_{re} { 1 \over H_{\phi \omega} -E-i \epsilon } \times 
\nonumber \\
&& V_{re}^* { 1 \over H_{K^* \bar K^*} -E-i \epsilon } + \cdots \biggr)V_{sb}^* \biggr] \Phi_G  = 0 \ ,
\nonumber \\
&& \left( H_{eff}(E)- { i \over 2} \Gamma(E) -E \right) 
\Phi_G  = 0 \  .
\label{eq:geometric}
\end{eqnarray}
Also using perturbation theory for $V_{re}$, we identify the imaginary
part of the first potential term, to order $V_{sb}^2$, as the partial
decay width to $K^* \bar K^*$, while the contribution (for an open
$\phi \omega$ channel) to the imaginary part from the second term, to
order $V_{sb}^2 V_{re}^2$, yields the partial decay width for $\phi
\omega$. The same analysis can be used for other sequential decays,
e.g. $G \rightarrow \pi \pi \rightarrow \omega \omega$.

In evaluating the string breaking and rearrangement potentials, we
truncate the sum over intermediate hadron states to the ground state
mesons in each channel $c$ and employ harmonic oscillator
wavefunctions, $\phi_{0}^{\alpha_c}$, having oscillator parameter
$\alpha_c$.  This yields the following separable potentials for string
breaking,
\begin{equation}
V_{sb} = v_{sb} | \phi_{0}^{\alpha_c} \rangle\langle \phi_{0}^{\alpha_c} | \ ,
\end{equation}
and rearrangement,
\begin{equation}
V_{re} = v_{re} | \phi_{0}^{\alpha_c} \rangle\langle \phi_{0}^{\alpha_c} | \  ,
\label{eq:rearrangement}
\end{equation}
and resulting partial widths
\begin{eqnarray}
\Gamma_{K^* \bar K^*}&=& 2 |v_{sb}|^2 Im\left[ g_{K^* \bar K^*}(E)\right]\ ,
\nonumber \\
\Gamma_{\phi \omega}&=& 2 |v_{sb}|^2 |v_{re}|^2 
Re\left[ g_{K^* \bar K^*}(E)\right] \times
\nonumber \\
&& Im\left[g_{\phi \omega}(E)\right] \, Re\left[g_{K^* \bar K^*}(E) \right] \ ,
\nonumber \\
g_c(E)&=&  \langle \phi_{0}^{\alpha_c} | {1 \over M_c+ {q^2 \over 2 \mu_c} -E - i \epsilon} | \phi_{0}^{\alpha_c} \rangle \ .
\end{eqnarray}
Here $M_c$ and $\mu_c$ are the threshold energy and reduced mass for
channel $c$.  Because the glueball mass is near the channel
thresholds, the real and imaginary parts of the channel Green
functions, $g_c(E)= a_c+ i\, b_c $, can be well approximated by
\begin{eqnarray}
a_c &\simeq& { \int_0^\infty { e^{-{\alpha_c}^2 q^2}   \over { q^2 \over 2 \mu_c } } q^2 d q  \over  \int_0^\infty 
{ e^{-{\alpha_c}^2 q^2} q^2 } d q } = 4 {\alpha_c} ^2 \mu_c
\\
b_c &\simeq&  4{\alpha_c}^3 \mu_c\sqrt{2 \pi \mu_c}\sqrt{E-M_c}
\end{eqnarray}
and $E=M_G$.
The partial decay widths are then
\begin{eqnarray}
\Gamma_{K^* \bar K^*}&=& 2 |v_{sb}|^2  b_{K^* \bar K^*} \ , 
\nonumber \\
\Gamma_{\phi \omega}&=& 2 |v_{sb}|^2 |v_{re}|^2   a_{K^* \bar K^*}^2 b_{\phi \omega} \ .
\end{eqnarray}

The only model dependent quantities are the potential strengths
$v_{re}$ and $v_{sb}$ but only one enters the ratio of the two partial
decay widths. Because the vector mesons have the same oscillator
parameter and the reduced masses $\mu_c$ are similar, while the
thresholds $M_c$ differ for ${K^* \bar K^*}$ and ${\phi \omega}$, the
ratio reduces to
\begin{eqnarray}
{ \Gamma_{\phi \omega} \over \Gamma_{K^* \bar K^*} }
&=& \left( v_{re} 4 \mu_{\phi \omega } \alpha_{\phi \omega }^2  \right)^2
\sqrt{ M_G- {M}_{\phi \omega } \over M_G  - {M}_{K^* \bar K^*} }  \  .
\label{eq:ratio}
\end{eqnarray}
The parameters used in our calculation are listed in Table
\ref{tablerescat}.  Notice that we did not compute the complete
geometric series in Eq. (\ref{eq:geometric}).  If the ratio,
Eq. (\ref{eq:ratio}), is large, we need to sum the full geometric
series in Eq. (\ref{eq:geometric}). However, the remaining terms of
the series contribute the same factor for the decay to $K^* \bar K^*$,
or to $\phi \omega$, and therefore the ratio is correct to all orders
in $V_{re}$.


\subsection{String breaking}

Because the ratio in Eq. (\ref{eq:ratio}) simplifies, we only list the
flavors directly produced with string breaking. Notice that the same
flavors are produced with a direct decay of the constituent
gluons. Let us suppose that there are two string breakings producing
two mesons.  We assume that each string breaking creates a
quark-antiquark pair in an approximately symmetric way, yielding an
SU(3) flavor singlet
\begin{equation}
u \bar u + d \bar d + s \bar s
\end{equation}
where we suppress spin and color notation. In the string breaking
picture, the quarks will be separated, each to one of the two produced
mesons. This is also necessary to ensure that each of the produced
mesons is a color singlet (equivalent to having a quark exchange
between the two flavor singlet sources and the two produced
mesons). Exchanging the first and third quarks (permutation operator
$P^{13}$) yields
\begin{eqnarray}
 &&P^{13} | (u \bar u + d \bar d + s \bar s) (u \bar u + d \bar d + s \bar s) \rangle
 \label{p13} 
\\
&=& | u \bar u u \bar u + d \bar d  d \bar d 
+ u \bar d d \bar u + d \bar u u \bar d  
\nonumber \\ &&
+ u \bar s s \bar u  + s \bar u u \bar s 
+ d \bar s s \bar d  + s \bar d d \bar s 
+ s \bar s s \bar s  
\rangle
\nonumber \\
&=& | {u \bar u + d \bar d \over \sqrt 2} {u \bar u + d \bar d \over \sqrt 2}
+  {u \bar u - d \bar d \over \sqrt 2} {u \bar u - d \bar d \over \sqrt 2}
\nonumber \\ &&
+ u \bar d d \bar u + d \bar u u \bar d  
+ u \bar s s \bar u  + s \bar u u \bar s 
+ d \bar s s \bar d  + s \bar d d \bar s 
+ s \bar s s \bar s  
\rangle
\nonumber
\end{eqnarray}
with a similar result for the exchange of the second and fourth
antiquarks ($P^{24}$).

Specializing to vector-vector production, nine different vector pairs
are produced and Eq. (\ref{p13}) becomes
\begin{eqnarray}
&=& | \omega \omega + \rho^0 \rho^0 + \rho^+ \rho^- + \rho^- \rho^+
\\ &&
+ K^{*+} K^{*-}+ K^{*-} K^{*+}+ K^{*0} \bar K^{*  0}+ \bar K^{* 0} K^{* 0}
+ \phi \phi
\rangle\; , 
\label{eq:vectors}
\nonumber
\end{eqnarray}
but no $\omega \phi$ from string breaking. This explains why the
off-diagonal Hamiltonian matrix element coupling this channel in
Eq. (\ref{eq:coupled}) is zero and that $\omega \phi$ can only be
produced by rearrangement.

\subsection {Rearrangement}

In evaluating $\Gamma_{\phi \omega}$ for the partial width ratio,
Eq. (\ref{eq:ratio}), the normalized $|K^* \bar K^* \rangle$ and $|
\omega \phi \rangle$ states are needed. The $ \omega \phi$ component
is
\begin{eqnarray}
|  \omega \phi \rangle &=& | { u \bar u + d \bar d \over \sqrt{2} } s \bar s \rangle \ , 
\end{eqnarray}
while $K^* \bar K^* $ is given by 
\begin{eqnarray}
|K^* \bar K^*  \rangle &=&\! \! |{ K^{*+} K^{*-}+ K^{*-} K^{*+}+ K^{*0} \bar K^{* 0} +  \bar K^{* 0} K^{*0} \over \sqrt 2 \sqrt 4 }\rangle 
\nonumber \\
&=& | { u \bar d d \bar u + d \bar u u \bar d  
+ u \bar s s \bar u  + s \bar u u \bar s   \over \sqrt 2 \sqrt 4 } \rangle \ ,
\end{eqnarray}
where the wavefunction normalization includes the meson-meson
exchange,
\begin{equation}
\langle K^* \bar K^*  | 1+ P^{13}P^{24} |K^* \bar K^*  \rangle =1 \ .
\end{equation}
Then the flavor rearrangement matrix element involving $P^{13}$
quark-quark exchange is
\begin{eqnarray}
\langle K^{*+ } K^{*-} | P^{13} | \phi \omega \rangle &=& { 1 \over 2 } \ .
\end{eqnarray}
Notice that $P^{24}$ antiquark-antiquark exchange produces exactly the
same result.  This also applies to the color and spin $\times$ space
rearrangement overlaps, so we only compute the $P^{13}$ overlaps and
include an additional factor of $2$ to account for antiquark exchange.

The color rearrangement overlap is 
\begin{eqnarray}
\langle 1 \ 1| P^{13} | 1 \ 1\rangle = { 1 \over 3 } \ .
\end{eqnarray}
For the space $\times$ spin matrix element, using the graphical rules
\cite{Ribeiro:1981fk, vanBeveren:1982sj}, a separable potential with
strength $v$ emerges
\begin{eqnarray}
\langle \phi_{0}^\alpha \ \phi_{0}^\alpha | P^{13} | \phi_{0}^\alpha \ \phi_{0}^\alpha \rangle = 
v | \phi_{0}^\alpha \rangle \langle \phi_{0}^\alpha | \  .
\end{eqnarray}
Hence, only the hyperfine interaction,  
\begin{equation}
V_{SS}{\ \lambda_i \cdot  \lambda_j  \over -16/3} 
\bs S_i \cdot \bs S_j  
\end{equation}
contributes to the space $\times$ spin rearrangement and the potential
$v$ is proportional to $V_{SS}\simeq 200 $ MeV obtained from
Ref. ~\cite{Bicudo:1995kq}.  To determine this constant we consider
the specific overlap where both $K^* \bar K^*$ and $\phi \omega$ have
a total spin 0. Coupling two vector mesons yields
\begin{eqnarray}
| 00 \rangle &=&  { |11\rangle |1-1\rangle - |10\rangle |10\rangle + |1 -1\rangle|11\rangle \over \sqrt{3} } 
\\
  &=& | { \uparrow  \uparrow \downarrow \downarrow + \downarrow \downarrow \uparrow  \uparrow \over \sqrt{3} } 
- { \uparrow  \downarrow \uparrow \downarrow + \uparrow  \downarrow \downarrow \uparrow + 
\downarrow \uparrow \uparrow \downarrow +\downarrow \uparrow \downarrow \uparrow \over 2 \sqrt{3} }\rangle
\nonumber 
\end{eqnarray}
so that
\begin{equation}
\langle 00 | P^{13} | 00 \rangle = -\frac{1}{2} \ .
\end{equation}
Only the intra-cluster contributions $V_{13}$, $V_{14}$, $V_{23}$,
$V_{24}$ of the hyperfine potential need to be considered
\cite{Bicudo:1989sj,Bicudo:1999kw}
since the inter-cluster ones are already included in the meson mass
calculation. Adding all intracluster contributions, the total space
$\times$ spin overlap contribution for this case is
\begin{equation}
-{ 3 \over 2 } V_{SS} | \phi_{0}^\alpha \rangle \langle \phi_{0}^\alpha | \  .
\end{equation}

\par
Then the resulting color $\times$ flavor $\times$ space $\times$ spin
overlap contribution for this case is
\begin{equation}
-{ 1 \over 2 } V_{SS} | \phi_{0}^\alpha \rangle \langle \phi_{0}^\alpha | \  ,
\end{equation}
yielding the strength and sign for the rearrangement potential
\begin{equation}
v_{re}=-{ 1 \over 2 } V_{SS} \ .
\end{equation}
The value of $\alpha$ is fixed by the rms radius, $<r^2>$, of the
corresponding meson.  For a Gaussian wavefunction
\begin{eqnarray}
<r^2> &=& \int_0^\infty dr \ r^4 e^{-r^2 / \alpha^2} \over \int_0^\infty dr \ r^2 e^{-r^2 / \alpha^2}.
\nonumber \\
&=& {3 \over 2} \alpha^2 \ .
\end{eqnarray}
For the pion $<r^2>^{1/2} \simeq$ 0.5 fm, but this is anomalously
large due to the light mass (in chiral perturbation theory it is
divergent in the chiral limit). We use 0.4 fm for kaons and 0.3 fm for
all vector mesons.  Converting to MeV$^{-1}$, Table \ref{tablerescat}
lists the oscillator parameters used for each channel.

\begin{table} [b]
\begin{tabular}{|c|cccccc|}
 \hline
 $c \rightarrow$  & $\phi\omega$ & $\omega\omega$ & $K^*\bar K^*$ & $\rho\rho$ & $K \bar K$ 
&$\pi\pi$ \\ \hline
$10^3 a_c ({\rm MeV^{-1}})$ & 2.74 &2.41 & 2.75 & 2.38 & 2.16 & 1.18 \\ 
$10^3 b_c ({\rm MeV^{-1}})$ & 0.57 & 2.35 &0.96 & 2.40 & 3.51 & 1.20 \\
$\alpha^{-1}_c ({\rm MeV})$ &  804 &  804 & 804 & 804 &603 &483 \\
$\mu_c ({\rm MeV})$ & 443 & 391 & 446 & 385 & 248 & 69\\
$M_{\rm c} ({\rm MeV})$ & 1802 & 1564 & 1784 & 1540 & 994 &  278\\
\hline
\end{tabular}
\caption{Rescattering parameters.}
\label{tablerescat}
\end{table}


\end{document}